\documentclass[prx,10pt,twocolumn,superscriptaddress]{revtex4-1}

\usepackage{graphicx}
\usepackage{rotating}
\usepackage{amssymb}
\usepackage{mathptmx}
\usepackage{diagbox}
\usepackage{multirow}
\usepackage{textgreek}
\usepackage{textcomp}
\newcommand{\RSD}{\sigma_{I_{\rm c}}/\left\langle I_{\rm c} \right\rangle}
\newcommand{\RSDf}{\sigma_{f}/\left\langle f \right\rangle}

\begin{document}

\newcommand{\hdblarrow}{H\makebox[0.9ex][l]{$\downdownarrows$}-}
\title{Improving wafer-scale Josephson junction resistance variation \\ in superconducting  quantum coherent circuits}

\author{J.M. Kreikebaum}
\altaffiliation{Author to whom correspondence should be addressed: jmkreikebaum@berkeley.edu}
\affiliation{Lawrence Berkeley National Laboratory, Berkeley, CA 94720, USA}
\affiliation{Department of Physics, University of California, Berkeley, CA 94720, USA}
\author{K.P. O'Brien}
\affiliation{Department of Electrical Engineering and Computer Science, Massachusetts Institute of Technology, Cambridge, MA 02139, USA}
\author{A. Morvan}
\affiliation{Lawrence Berkeley National Laboratory, Berkeley, CA 94720, USA}
\author{I. Siddiqi}
\affiliation{Lawrence Berkeley National Laboratory, Berkeley, CA 94720, USA}
\affiliation{Department of Physics, University of California, Berkeley, CA 94720, USA}

\date{\today}

\begin{abstract}
Quantum bits, or qubits, are an example of coherent circuits envisioned for next-generation computers and detectors. A robust superconducting qubit with a coherent lifetime of $O$(100 \textmu s) is the transmon: a Josephson junction functioning as a non-linear inductor shunted with a capacitor to form an anharmonic oscillator.  In a complex device with many such transmons, precise control over each qubit frequency is often required, and thus variations of the junction area and tunnel barrier thickness must be sufficiently minimized to achieve optimal performance while avoiding spectral overlap between neighboring circuits. Simply transplanting our recipe optimized for single, stand-alone devices to wafer-scale (producing 64, 1x1 cm dies from a 150 mm wafer) initially resulted in global drifts in room-temperature tunneling resistance of $\pm$ 30\%. Inferring a critical current $I_{\rm c}$ variation from this resistance distribution, we present an optimized process developed from a systematic 38 wafer study that results in $<$ 3.5\% relative standard deviation (RSD) in critical current ($\equiv \RSD$) for 3000 Josephson junctions (both single-junctions and asymmetric SQUIDs) across an area of 49 cm$^2$. Looking within a 1x1 cm moving window across the substrate gives an estimate of the variation characteristic of a given qubit chip. Our best process, utilizing ultrasonically assisted development, uniform ashing, and dynamic oxidation has shown $\RSD$ = 1.8\% within 1x1 cm, on average, with a few 1x1 cm areas having $\RSD$ $<$ 1.0\% (equivalent to $\RSDf$ $<$ 0.5\%). Such stability would drastically improve the yield of multi-junction chips with strict critical current requirements.
\end{abstract}

\maketitle

\section{Introduction}
Josephson junctions, fabricated by isolating two superconductors with a thin insulating barrier, are the core circuit element for superconducting solid state quantum coherent devices. When shunted with a capacitor, the non-linear inductance from the junction forms an anharmonic oscillator making energy levels individually addressable \cite{Koch2007}. Precise control over junction properties is crucial for state-of-the-art devices such as: quantum processors utilizing the cross-resonance gate \cite{Tripathi2019}, single microwave photon detectors based on ensembles of identical qubits \cite{Royer2018}, and travelling wave amplifiers where variations in nominally identical junctions lead to unwanted impedance variations \cite{Macklin2015}. Therefore, in this work we specifically focus on the reproducibility of shadow-evaporated sub-micron Al/AlO$_{x}$/Al Josephson junctions common to nearly all current qubits \cite{Krantz2019}.

The critical current, $I_{\rm c}$, of a Josephson junction, inversely proportional to its inductance, is tuned by either varying the critical current density, $J_{\rm c}$, or the junction area. The former involves modifying the tunnel barrier thickness via the oxidation time or pressure when using a thermally grown barrier. Our wafer-scale fabrication process produces 64, 1 cm$^2$ dies from a 150 mm wafer \textemdash~ the maximum size accommodated by our evaporator. The junctions are located within the central $\approx$ 49 cm$^2$ of the die array and thus high uniformity is desired over this length scale. Previous works describe two types of Josephson tunnel junctions: large junctions, $I_{\rm c}$ $O$(\textmu A), typically realized with a Nb/AlO$_{x}$/Nb trilayer process suitable for superconducting digital electronics or microwave amplifiers; small junctions, $I_{\rm c}$ $O$(nA), typically realized with Al/AlO$_{x}$/Al suitable for qubits. Regarding the former, 2-4\% intrachip variations have been reported \cite{Bumble2009} and  $\approx$ 15\% variation is observed across a wafer \cite{Ketchen1991,Lotkhov2018}; a notable exception is \cite{Tolpygo2015} where 8.2\% and 2.9\% variation in resistance is reported for 300 nm and 800 nm diameter junctions, respectively, across a 200 mm wafer.  Junctions with sizes ranging from 0.015 to 3.27 \textmu m$^2$ mentioned in \cite{Krantz2010} had variations of 2.3\% on 39 mm$^2$ chips. For qubits, it is advantageous to reduce the physical size of the junction to minimize the inclusion of noisy two level defects \cite{Nugroho2013}. Authors fabricating deep sub-micron junctions typically report fluctuations of $\approx$ 5\% within chips smaller than 50 mm$^2$ \cite{IBMMM2019}, 3.5\% within a few mm$^2$ \cite{Pop2012},  and fluctuations of 2-3\% for 0.04 \textmu m$^2$ junctions patterned with hard masks across 50 mm wafers \cite{Niedzielski2019}. 

In this work, we strive to further improve this absolute level of resistance variation, and to realize it over a larger substrate in order to increase the yield of functional multi-qubit chips which have tight tolerances on qubit frequency. Furthermore, we investigated designs where a SQUID replaces a single junction and the magnetic flux-tunability of the circuit inductance is limited by introducing asymmetry in the SQUID junction areas ($\geq$ 5:1) to reduce the susceptibility to flux noise \cite{Koch2007, Hutchings2017}. As such, we produced small junctions over a range of areas spanning 0.0036 to 0.013 \textmu m$^2$. It is important to note that in such SQUIDs, the smaller junction only affects the tuning range so we focus on tight control over the critical current of the larger junction.

\begin{figure*}[ht]
\centering
\includegraphics[width=\textwidth, keepaspectratio]{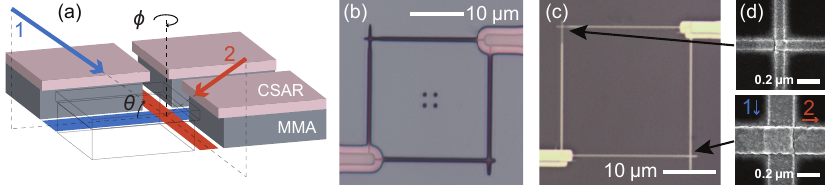}
\caption{Device geometry for the asymmetric SQUID used in this study. (a) Sketch of the resist stack for ``Manhattan Style'' junctions. Developed features are designed to be deeper than their width to allow metal to reach the substrate when evaporated parallel to a given channel, but block metal in orthogonal channels. Thermal oxidation of layer 1 occurs before rotating the substrate and depositing layer 2. A third layer, rotated by $\phi$ = 180$^{\circ}$ relative to layer 2 (not shown) is required to form the second SQUID junction. A high sensitivity resist (MMA) results in a undercut of the high resolution top layer (CSAR) to improve liftoff quality. (b) Micrograph of the developed resist stack. (c) Micrograph of the final SQUID structure. (d) Scanning electron microscope images of the two Josephson junctions forming the 8:1 asymmetric SQUID on wafer 37.}
\label{fig1}
\end{figure*}

\section{Methods and Observations}
\begin{figure*}[ht]
\centering
\includegraphics[width=\textwidth, keepaspectratio]{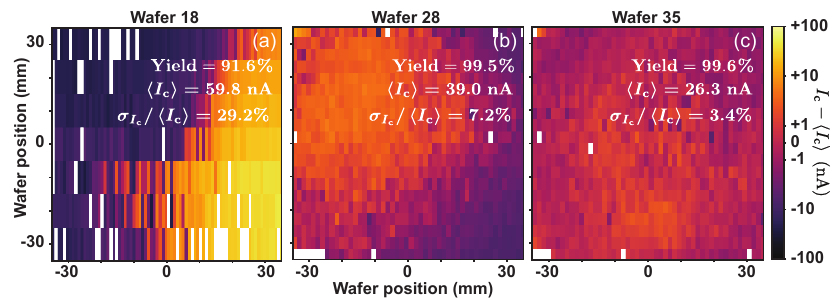}
\caption{Select wafer-scale resistance variation data for 6:1 asymmetric SQUIDs. Shown here are improvements in uniformity over the course of this study with inferred $I_{\rm c}$ plotted as the difference from the mean of each wafer with a common z-axis. White cells indicate a junction that was open or shorted. (a) The first wafer-scale statistics acquired (wafer 18). (b) First wafer probed with the automated probe station and after implementation of dynamic oxidation and attenuated ultrasonic MMA development (wafer 28). (c) Current uniformity after implementing improved ashing, slower evaporation rate, and larger junctions with lower $J_{\rm c}$.  Yield is calculated excluding failures due to patterning junctions in the resist edgebead.}
\label{fig2}
\end{figure*}

For this study, both 100 and 150 mm wafers were used. Junctions were fabricated using the bridge-free ``Manhattan Style" \cite{Costache2012,Potts2001} on $>$ 8000 \textOmega-cm intrinsic (100) Si using e-beam lithography, see Fig.~\ref{fig1}. Bridgeless junctions have an advantage over bridged designs, such as Dolan style \cite{Dolan1977}, that the junction area is independent of resist thickness. Layouts were generated in python with GDSpy \cite{GDSpyRepo}, proximity effect corrected with Beamer from GenISys, and exposed with 100 keV electrons in a Raith Electron Beam Pattern Generator (EBPG) 5150. The EBPG is housed in an enclosure made by MCRT within a class 100 cleanroom. The enclosure re-filters the air to at least class 10 and stabilizes temperatures to $\pm$ 0.05 $^{\circ}$C over month-scale time frames. A Spicer Consulting SC24 provides active 3-axis magnetic field cancellation from DC-13 kHz, measured at a single point next to the e-beam column. The environmental stability of the setup, combined with the Raith EBPG 5150 self-calibration protocol, provides highly reproducible lithography. Once exposed, samples are developed and subsequently coated with e-beam evaporated Al in a Plassys MEB550s with a base pressure of $3 \times 10^{-8}$ mbar. After liftoff, junctions were individually probed to measure their room temperature resistance from which $I_{\rm c}$ can be inferred using the Ambegaokar-Baratoff formula \cite{Ambegaokar1963}. These values can be converted into a qubit frequency using an estimate of the shunt capacitance. Initially, wafers were probed by hand but later, a Micromanipulator P200L semi-automatic probe station was used for the last 11 wafers to gather statistics on a larger number of junctions. Plots highlighting improvements made during this study can be found in Fig.~\ref{fig2}.

\subsection{Resist/Exposure}
The resist bi-layer was spun with a Laurell Technologies WS-650-23B spin coater. MicroChem MMA-EL13 (copolymer in ethyl lactate) was used as the high sensitivity bottom undercut layer for all wafers. Zeon Corp. ZEP 520A-7, MicroChem 950k PMMA A4, and AllResist GmbH AR-P 6200.9 (CSAR) were all tested as the high resolution upper layer. It was found that the small ($\approx$ 20 mm diameter) vent hole in the top of the spin coater had to be covered to create a uniform spin of the MMA, which was unnecessary for the CSAR and ZEP likely because of the differences in viscosity of anisole and ethyl lactate. We initially had difficulties spinning defect free CSAR on MMA, behavior which was not observed with ZEP. This issue was solved after the resist was degassed by opening the lid and letting it sit for 2 hours allowing the pressure and humidity in the bottle to equilibrate with ambient conditions. CSAR was ultimately selected as the resist of choice over PMMA because of the flexibility it offered having (mostly) orthogonal development chemistry to MMA and over ZEP because of its lower cost. For our developers, described below, MMA and CSAR had an optimal dose of 180 and 1100 \textmu C/cm$^2$ respectively. We note that partial clearing of CSAR in MMA developer was observed for doses above 1100 \textmu C/cm$^2$ when immersed for extended times.

Proximity effect correction (PEC) in Beamer was first optimized by observing the uniformity (or lack) of residual undercut as the MMA provides a sensitive indicator of long range substrate backscattering compensation. The software's 3D-Edge mode of 3D PEC was chosen due to its ability to simultaneously proximity effect correct both resist layers which require different doses and a default point spread function (PSF): 500 nm PMMA on Si at 100 keV (Z-Position: 0.325) was used initially. Before the addition of short range corrections to this PSF, we had low yield of sub 100 nm features with CSAR which we did not observe with ZEP. The short range corrections that were added to improve yield were: an effective short range blur FWHM of 50 nm, a short range separation value of 5 \textmu m, and a mid-range activation threshold of 2\%. A 200 pA beam and 200 \textmu m aperture (calculated spot size = 2 nm) was used with a 1 nm beam step size to ensure that designed area variations on the order of a few nm were reproduced. Backscatter dosing from the probe pads (which are not written on device wafers) were written 130 \textmu m away ($\sim$ 4x the backscattering parameter for 100 keV electrons on Si) to ensure test wafers created junctions equivalent to device wafers.

SEM observations of as-evaporated junctions showed worse line edge roughness (LER) on the second evaporation compared to the first (see Fig.~\ref{fig1}). Our theory is that Al deposited on the sidewall of the CSAR in the first evaporation introduces additional LER for subsequent evaporations. A trilayer resist (MMA/CSAR/MMA) was briefly considered in an attempt to reduce this effect utilizing the top layer of MMA to shield the CSAR during off-axis evaporations. We did observe an improvement in LER, but since it did not reduce global $I_{\rm c}$ variations, it was abandoned due to its added complexity and the additional forward beam scattering from the top MMA would result in increased developed linewidths \cite{Gorelick2010}, limiting achievable SQUID asymmetry ratios.

\subsection{Development}
Cold development with manual agitation (or ultrasonication for wafer 36) was used for CSAR and ZEP. A Thermo Scientific PC200 immersion circulator filled with 50:50 H$_{2}$O: Propylene Glycol was used to chill N-amyl acetate (NAA) baths to 0 $\pm$ 0.02 $^{\circ}$C. NAA from Zeon corp. (ZED-N50) was used initially and AllResist GmbH AR 600-546 was used after wafer 26. No difference was noted between these nominally identical developers. The MMA was developed at room temperature and puddle development was briefly considered, but led to many CSAR constrictions so was abandoned in favor of immersion development on PTFE wafer holders. Initially IPA:MIBK was used to develop the MMA but we observed many open junctions due to small resist bridges constricting the CSAR near the junction, especially for $<$ 0.01 \textmu m$^2$ junctions. Our hypothesis was that swollen, gel-like MMA removed by the developer \cite{Papanu1989} was the cause of these constrictions. Studies with PMMA (which has much higher molecular weight than MMA), showed that the co-solvent IPA:H$_{2}$O was a superior developer, resulting in reduced swelling and the addition of sonication was shown to increase the rate at which developed resist is removed \cite{Rooks2002, Mohsin1988, Hasko2000, Yasin2002}. Although the switch of developer alone did not drastically improve small junction yield, the addition of sonication did. Care had to be taken to attenuate the ultrasonication power to prevent collapse of the CSAR overhang which was accomplished by using the lowest bath power and, crucially, lining the bath with a polyurethane/vinyl sound absorbing foam, leaving the central 1x1 cm open to allow some power transmission.

After development, oxygen plasma ashing of the newly opened channels is performed. We used a Plasma Etch PE-50 with a 50 kHz pure oxygen plasma (80 s, $\approx$ 500 mbar, $\approx$ 60 W). It was found that large, non-radially symmetric $I_{\rm c}$ gradients were reduced and made more radially symmetric by splitting a single ashing step into four, 20 s steps with 90 $^{\circ}$ substrate rotations between steps. In an attempt to further improve the ashing uniformity, the sample was rotated four times in each corner of the chamber, for a total of 16 x 5 s ashes. This resulted in the best wafer-scale statistics at the time: $\RSD = 3.5\%$ for single junctions across 49 cm$^2$. Eliminating ashing resulted in worse $\RSD$ but also a 2x reduction in $J_{\rm c}$, strong evidence that residual organics have an effect on tunnel barrier properties \cite{Pop2012, Koppinen2007, Zeng2015b}.

\begin{figure*}[ht]
\includegraphics[width=\textwidth, keepaspectratio]{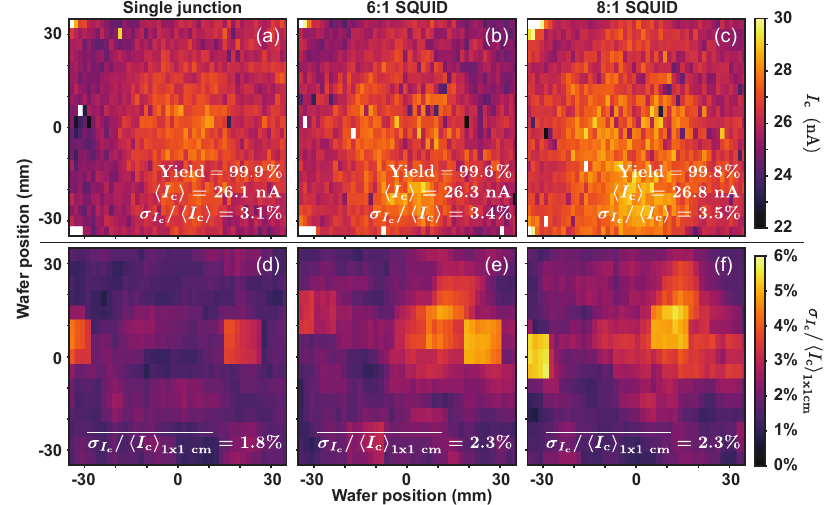}
\centering
\caption{Wafer map of Josephson junction critical currents observed for wafer 35. (a, b, c) $I_{\rm c}$ values with white cells indicating a junction that was open or shorted. (d, e, f) $\RSD$ for the 27 junctions of a single design within a 1x1 cm moving window. (a, d) Fixed frequency single junction. (b, e) 6:1 asymmetric SQUID. (c, f) 8:1 asymmetric SQUID.}
\label{fig3}
\end{figure*}

After implementing 16x ashing, the dominant source of non-uniformity was found to be junction area variations which showed approximately radial dependence. First, the effect was reduced simply by increasing the junction area (and decreasing $J_{\rm c}$ to keep $I_{\rm c}$ constant). To test if this was introduced during development in NAA, manual agitation was replaced by ultrasonication for wafer 36 due contrast improvements seen in \cite{Tobing2012} and assumed higher uniformity. However, this showed no improvement and a $\sim$ 1 cm$^2$ patch of abnormally low $J_{\rm c}$ on the wafer caused an overall $\RSD$ degradation. Pinpointing the cause of, and a solution to, the area fluctuations is the path towards better wafer-scale uniformity for this process. To this end, a hard mask process would be helpful as it should be more robust during evaporation and diagnostic post-development SEM imaging.

\subsection{Evaporation and Oxidation}
Motivated by the hypothesis that high energy electrons and UV radiation released during the evaporation could warp or distort the resist non-uniformly and produce the observed area fluctuations, a deposition rate of 3 nm/s was used for the majority of this study. However after other process optimizations, better uniformity was observed using a rate of 0.3 nm/s. Lower deposition rates provide more time for a growing film and substrate to thermalize, forming smaller grains \cite{Hentzell1984}. Since the tunnel barrier thickness is not uniform grain to grain or at grain boundaries \cite{Fritz2018, Zeng2015, Nik2016}, we hypothesize that more grains per junction results in better averaging of the effective barrier thickness, improving site to site uniformity. To investigate this, cross section TEM analysis of junctions fabricated with the two deposition rates is ongoing \cite{Liu2020}. Dynamic and static oxidations were also A/B tested. In a static oxidation, the chamber is filled with oxygen (in our case 95\%/5\% Ar/O) to a set pressure and then evacuated after a set time. In a dynamic oxidation, gas is continuously introduced and pumped out with rates balanced such that the pressures are the same as the static oxidation case. Interestingly, we found dynamic oxidation produced a lower $J_{\rm c}$ and since it provided better uniformity, it was used for the remainder of the study. 

\section{Results}

\begin {table*}[ht]
\tiny
\resizebox{0.9\textwidth}{!}{
 \begin{tabular}{|c c||c c c c c c c c c c c|} 
 \hline
 \multicolumn{2}{|c||}{\backslashbox{Parameter}{Wafer}} & 28 & 29 & 30 & 31 & 32 & 33 & 34 & 35 & 36 & 37 & 38\\ [0.5ex] 
 \hline\hline
  \multicolumn{2}{|c||}{\multirow{2}{*}{oxidation}} & {\multirow{2}{*}{dyn}} & {\multirow{2}{*}{stat}} & {\multirow{2}{*}{dyn}} & {\multirow{2}{*}{dyn}} & {\multirow{2}{*}{dyn}} & {\multirow{2}{*}{dyn}} & dyn & dyn & dyn & dyn & dyn\\
  & &  &  &  &  &  &  & 2P, 2t & 2P, 2t & 2P, 2t & 2P, 2t & 2P, 2t\\
 \hline
 \multicolumn{2}{|c||}{ashing} & 4x & 4x & 16x & 16x & none & 4x & 4x & 16x & 16x & 16x & 16x\\ 
 \hline
 \multicolumn{2}{|c||}{evaporation rate (nm/s)} & 3.0 & 3.0 & 3.0 & 3.0 & 3.0 & 0.3 & 0.3 & 0.3 & 0.3 & 0.3 & 0.3\\ 
 \hline
 \multicolumn{2}{|c||}{fresh NAA} & y & n & n & y & y & y & y & y & y & y & y\\ 
 \hline
 \multicolumn{2}{|c||}{agitation during MMA dev} & y & n & n & y & y & y & y & y & y & y & y\\
 \hline
 \multicolumn{2}{|c||}{CSAR dev ultrasonication} & n & n & n & n & n & n & n & n & y & n & n\\
 \hline
 \multicolumn{2}{|c||}{single junction design size} & A & A & A & A & A & A & 1.06A & 1.06A & 1.06A & 1.4A & 1.4A\\ 
  \hline
  \multicolumn{2}{|c||}{single JJ $\overline{I_{\rm c}}$ (nA)} & 65.6 & 94.7 & 63.1 & 64.2 & 30.8 & 55.3 & 34.1 & 26.1 & 26.3 & 32.6 & 33.2 \\ 
  \hline
 \multicolumn{2}{|c||}{6:1 SQ junction design size} & A & A & A & A & A & A & 1.7A & 2.2A & 2.2A & 2.9A & 2.9A\\
 \hline
  \multicolumn{2}{|c||}{6:1 SQ $\overline{I_{\rm c}}$ (nA)} & 39.0 & 47.0 & 35.0 & 37.6 & 20.5 & 29.0 & 27.7 & 26.3 & 27.0 & 34.5 & 34.7\\
 \hline
 \multirow{3}{*}{$\RSD$ (\%)}& single JJ & 6.5 & 9.1 & 3.5 & 5.2 & 10.8 & 3.6 & 4.4 & 3.1 & 3.7 & 3.8 & 3.1\\
    & 6:1 SQ & 7.2 & 9.6 & 9.2 & 6.7 & 7.8 & 7.5 & 4.9 & 3.4 & 5.0 & 4.1 & 4.9\\
    & 8:1 SQ & 7.5 & 8.4 & 8.0 & 6.9 & 8.2 & 7.4 & 4.9 & 3.5 & 5.1 & 4.3 & 4.7\\
\hline
\multirow{3}{*}{$\overline{\sigma_{I_{\rm c}}/\left\langle I_{\rm c} \right\rangle_{\rm 1x1 cm}}$ (\%)}& single JJ & 1.8 & 2.8 & 2.3 & 3.5 & 6.7 & 1.7 & 2.5 & 1.8 & 2.0 & 1.9 & 2.1\\
    & 6:1 SQ & 2.2 & 5.6 & 5.2 & 5.3 & 5.3 & 5.1 & 2.8 & 2.3 & 2.5 & 1.9 & 2.0\\
    & 8:1 SQ & 2.3 & 5.1 & 5.0 & 4.9 & 5.1 & 5.1 & 2.8 & 2.3 & 2.4 & 2.1 & 1.9\\
 \hline
\end{tabular}}\\
\normalsize
\caption{\label{tab:summary} Summary of modified process variables and uniformity results for the 11 wafers measured using automatic probing. The aluminum crucible was refilled after wafer 34. 2P, 2t refers to double oxidation pressure and time compared to unspecified cases. Agitation during MMA development is a gentle manual agitation of the wafer in the ultrasonic bath. Junction design size specifies nominal relative junction areas, useful when comparing average $I_{\rm c}$ between wafers. Wafer-to-wafer repeatability can be evaluated by comparing wafers 35/36 and 37/38.}
\end{table*}

Wafers (which each had 1000 fixed frequency junctions, 1000 6:1 SQUIDs, and 1000 8:1 SQUIDs patterned in alternating rows of 50) made after delivery of the automated probe station  are summarized in Table~\ref{tab:summary}. The full recipe yielding the highest uniformity can be found in the supplementary material \cite{SupMat} and the junction properties from wafer 35 are plotted in Fig.~\ref{fig3}. 

\subsection{Qubit Coherence and Frequency Predictability}
Many measurements are still needed to rigorously correlate the observed improvements in junction uniformity with ultimate device performance. Nonetheless we describe here example measurements of two co-fabricated 8-qubit quantum processors. The 8 fixed-frequency transmon qubits on each chip had a mean target frequency of 5.6 GHz with detunings between neighbors optimized for the cross-resonance gate \cite{Tripathi2019}. Fabricating 64 chips on a 150 mm Si wafer and binning the 64 junctions of each size across the wafer, we find an average $\RSD$ = 6.9\%, a 3.8x improvement over an 8-qubit ring wafer made using a process similar to wafer 29 (where MIBK was used instead of H$_{2}$O for MMA development). With this narrower distribution of critical currents, we found 3/64 chips had optimal qubit frequencies, consistent with numerical estimates of chip yield given the measured $\RSD$. We hypothesize that the remaining $\sim$ 2x discrepancy in $\RSD$ between test wafers 35, 37 and 38 and the latest device wafer may be explained by the additional round of lithography that device wafers require after junction deposition (including resist baking and ion-milling) to define the low-loss junction-capacitor interconnects \cite{Dunsworth2017} or the different substrate surface between test and device wafers (RIE etched vs polished Si). The chips were wirebonded in two designs of Cu boxes and tested in separate dilution refrigerators. Coherence measurements and frequency predictability are summarized in Table~\ref{tab:coh}. Given the long lifetimes measured on sample \#1, we conclude that the fabrication modifications made to improve uniformity do not come at the expense of qubit coherence. See the supplementary material for discussion of the observed offsets between probing estimates and cryogenic measurements, the suppressed coherence of sample \#2, and data on individual qubits.

\begin {table}[h]
\footnotesize
\centering
\resizebox{0.8\linewidth}{!}{
 \begin{tabular}{|c|c|}
 \hline
 Sample \#1: & average of 8 qubits \\
 \hline
 f$_{01}$ (probing est.) (GHz) & 5.803 \\ 
  \hline
 f$_{01}$ (difference at 8 mK) & -2.62\% $\pm$ 0.50\% \\ 
  \hline
 T$_{1}$ (\textmu s) & 104\\ 
  \hline
 T$_{2}^{*}$ (\textmu s) & 60 \\ 
  \hline
 T$_{2Echo}$ (\textmu s) & 107 \\ 
  \hline
 \hline
 Sample \#2: & average of 8 qubits \\
 \hline
 f$_{01}$ (probing est.) (GHz) & 5.651 \\ 
  \hline
 f$_{01}$ (difference at 14 mK) & 0.38\% $\pm$ 0.40\%\\ 
  \hline
 T$_{1}$ (\textmu s) & 45 \\ 
  \hline
 T$_{2}^{*}$ (\textmu s) & 31 \\ 
  \hline
 T$_{2Echo}$ (\textmu s) & 49 \\ 
  \hline
\end{tabular}}
\normalsize
\caption {\label{tab:coh}Average qubit properties from a device wafer fabricated with the high-uniformity junction recipe of wafers 35, 37, and 38. Frequency estimates from room temperature junction resistance and simulated capacitance show good agreement with actual qubit frequencies, especially when considering detunings between neighbors. Coherences are quoted as the average value of the 8 qubits per chip. Data on individual qubits can be found in the supplementary material.}
\end{table}

\section{Conclusions}
Motivated by the challenging task of maintaining high Josephson junction uniformity when scaling quantum coherent circuit fabrication beyond a few qubits, we undertook a systematic study to identify and rectify sources of $I_{\rm c}$ variation. We have developed a process which has shown a $\RSD$ as low as 3.1\% over 49 cm$^2$ for single junctions. Looking within a chip sized 1 cm$^2$ window to remove global drift, an average $\RSD$ = 1.8\% was measured with some areas $<$1.0\%. To accomplish this, a reliable resist stack was found by changing proximity effect correction parameters and studying different development strategies, of which ultrasonication played a key role in producing high yield structures. Large gradients introduced by non-uniform ashing were mitigated by adding substrate rotations into that process, which may not be necessary with a more uniform asher. Slower evaporation rates and dynamic oxidations were then shown to further improve uniformity. Current levels of uniformity should be improved by minimizing the observed junction area fluctuations, whose origin is not currently understood. However, since $\RSD$ within chip sized areas is small, detunings between qubits on a single chip can be accurately set and the non-zero global $I_{\rm c}$ drift can be used to target absolute frequencies; a useful capability as tolerances become tighter for quantum processors and microwave photon detectors growing in complexity, size, and qubit number.

\begin{acknowledgments}
We thank I. Ene and W. P. Livingston for their assistance with code development, B. K. Mitchell and X. Liu for useful discussions, and Bleximo Corp. for designing and assembling a sample box. Work was supported by Samsung Electronics Co., Ltd. and the U.S. Department of Energy, Office of Science, Basic Energy Sciences, Materials Sciences and Engineering Division under Contract No. DE-AC02-05-CH11231 within the High-Coherence Multilayer Superconducting Structures for Large Scale Qubit Integration and Photonic Transduction program (QISLBNL).
\end{acknowledgments}

\end{document}

\title{Supplementary material for: Improving wafer-scale Josephson junction resistance variation in superconducting  quantum coherent circuits}

\author{J.M. Kreikebaum}
\altaffiliation{Author to whom correspondence should be addressed: jmkreikebaum@berkeley.edu}
\affiliation{Lawrence Berkeley National Laboratory, Berkeley, CA 94720, USA}
\affiliation{Department of Physics, University of California, Berkeley, CA 94720, USA}
\author{K.P. O'Brien}
\affiliation{Department of Electrical Engineering and Computer Science, Massachusetts Institute of Technology, Cambridge, MA 02139, USA}
\author{A. Morvan}
\affiliation{Lawrence Berkeley National Laboratory, Berkeley, CA 94720, USA}
\author{I. Siddiqi}
\affiliation{Lawrence Berkeley National Laboratory, Berkeley, CA 94720, USA}
\affiliation{Department of Physics, University of California, Berkeley, CA 94720, USA}

\date{\today}

\maketitle

\section{Detailed recipe (used for wafer 35, 37, and 38)}
\label{sec:recipe}
\begin{itemize}
  \item Clean substrate: 30 s in 5:1 BOE, DI rinse, IPA rinse, spin dry.
  \item 60 s bake 200$^{\circ}$ C, cool 60 s with N$_{2}$ gun 
  \item Pour MMA-EL13, spin (vent hole closed) 90 s @ 1 krpm, 1 krpm/s (film $\approx$ 500 nm)
  \item 90 s bake 150$^{\circ}$ C
  \item Pour AR-P 6200.9, spin (vent hole open) 60 s @3 krpm, 1 krpm/s (film $\approx$ 150 nm)
  \item 60 s bake 150$^{\circ}$ C
  \item Expose in Raith EBPG 5150 at 100 keV, base dose: 180/1100 \textmu C/cm$^2$ for MMA/CSAR
  \item Develop AR-P: 60 s n-amyl acetate at 0$^{\circ}$ C with manual agitation
  \item Stop development with 10 s dip in IPA bath
  \item Develop MMA: 75 s in 3:1 IPA:H$_{2}$O with attenuated ultrasonication
  \item Stop development with 10 s dip in IPA bath, N$_{2}$ dry
  \item Ash: 60 W, 500 mbar, 5 s x 16 orientations (90$^{\circ}$ rotations in each corner of asher)
  \item Evaporate: 
    \begin{itemize}
     \item Pump for 24 hours to $\approx$ $4 \times 10^{-8}$ mbar; getter with Ti (3 min, 0.2 nm/s)
     \item Pressure now $\approx$ $2 \times 10^{-8}$ mbar, evaporate Al (30 nm, 0.3 nm/s, $\theta,\phi$ = 45, 0)
     \item 1 rpm $\phi$ rotation: 10 min cooldown \& 20 min, 20 mbar dynamic oxidation
     \item Getter with Ti (3 min, 0.2 nm/s), pressure now $\approx$ $4 \times 10^{-8}$ mbar 
     \item Evaporate Al (30 nm \& 40 nm, 0.3 nm/s $\theta,\phi_{1},\phi_{2}$ = 45, -90, 90)
    \end{itemize}
  \item Liftoff 2 hours in acetone at 67$^{\circ}$C, placement into fresh acetone and sonicated at low power for 2 min with removal into IPA stream and N$_{2}$ dry
  \item Ash: 3 min, 80 W, 500 mbar, then probe and SEM.
  \\
\end{itemize}

\section{Qubit Coherence and Frequency Predictability}

\begin{figure}[ht]
  \caption{The multi-qubit chip architecture used for this study.}
  \centering
    \includegraphics[width=0.5\textwidth]{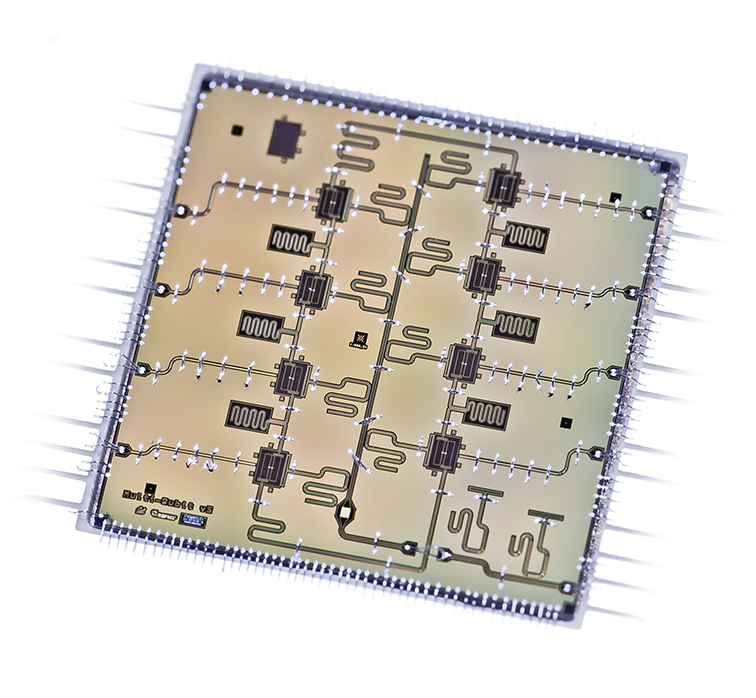}
\end{figure}
The fixed-frequency transmon qubits ($E_{J}/E_{C} = 70$) measured in this study are coupled to nearest neighbors forming a ring topology. Qubit state is dispersively readout with frequency multiplexed $\lambda/4$ CPW resonators ($\kappa_{ext} \approx 1$ MHz) coupled to a common $\lambda/2$ CPW resonator acting as a Purcell filter to allow fast readout without excess qubit energy relaxation into the bus. In addition to a control line for each qubit, these structures are defined with Cl based inductively coupled reactive ion etching of sputtered Nb on $>$ 8000 \textOmega-cm intrinsic (100) Si. After resist stripping in Microposit Remover 1165 at 80$^{\circ}$ C for 1 hour, junctions were added using the recipe above. The final round of lithography opens a window above the junction-capacitor overlap region, which was then ion-milled at 400 V for 6 min (with a duty cycle of 2 s on and 20 s off to prevent excessive substrate heating/resist cross-linking) and subsequently covered with a 200 nm normal angle Al deposition serving as a low-loss junction-capacitor galvanic interconnect \cite{Dunsworth2017}. Two test junction sites within each die were also fabricated for a total of 640 junctions on the wafer. We probed the room temperature resistance of these test sites approximately every 48 hours for a month with the wafer continuously exposed to air to investigate any aging of the junctions. Fitting the normalized change in junction resistance with an exponential function, we find a fit amplitude of 5.5 $\pm$ 0.5\% and a time constant of 132 $\pm$ 37 hours. In order to minimize the physical damage that occurs when the probe tips touch the surface, the actual qubit sites were only probed after the wafer aged into the steady state. In order to remove all resist residue, chips are cleaned in Microposit Remover 1165 at 80$^{\circ}$ C for 12 hours. This cleaning treatment induces additional aging of 4.3 $\pm$ 0.8\% in $I_{c}$ (with an exponential rise to this new steady state in $\approx$ 48 hours) so a few candidate dies are cleaned and the exact chip is chosen with post-1165 probing data. Chips are then wirebonded in a copper cryopackage and mounted to the mixing chamber of a dilution refrigerator. Additional wirebonds are used on chip to suppress unwanted slotline modes.

\begin {table*}[ht]
% \footnotesize
\resizebox{\textwidth}{!}{
 \begin{tabular}{|c|c c c c c c c c c|}
 \hline
 Sample \#1: & average & QB1 & QB2 & QB3 & QB4 & QB5 & QB6 & QB7 & QB8\\
 \hline
 f$_{01}$ (probing est.) (GHz) & 5.803 & 5.748 & 5.607 & 5.674 & 5.798 & 5.860 & 5.984 & 5.885 & 5.866 \\ 
  \hline
 f$_{01}$ (difference at 8 mK) & -2.62\% $\pm$ 0.50\% & -1.82\% & -2.68\% & -3.00\% & -3.23\% & -2.96\% & -2.37\% & -2.05\% & -2.81\% \\ 
  \hline
 T$_{1}$ (\textmu s) & 104 & 114 $\pm$ 20 & 127 $\pm$ 15 & 115 $\pm$ 24 & 133 $\pm$ 13 & 115 $\pm$ 17 &  75 $\pm$ 13 & 104 $\pm$ 12 & 49 $\pm$ 12\\ 
  \hline
 T$_{2}^{*}$ (\textmu s) & 60 & 30 $\pm$ 8 & 100 $\pm$ 9 & 98 $\pm$ 16 & 61 $\pm$ 21 & 49 $\pm$ 14 & 15 $\pm$ 4 & 81 $\pm$ 8 & 47 $\pm$ 9\\ 
   \hline
 T$_{2Echo}$ (\textmu s) & 107 & 77 $\pm$ 18 & 128 $\pm$ 10 & 114 $\pm$ 14 & 132 $\pm$ 8 & 100 $\pm$ 31 & 82 $\pm$ 10 & 106 $\pm$ 11 & 60 $\pm$ 11 \\ 
  \hline
 \hline
 Sample \#2: & average & QB1 & QB2 & QB3 & QB4 & QB5 & QB6 & QB7 & QB8\\
 \hline
 f$_{01}$ (probing est.) (GHz) & 5.651 & 5.588 & 5.411 & 5.507 & 5.584 & 5.698 & 5.834 & 5.881 & 5.703 \\ 
  \hline
 f$_{01}$ (difference at 14 mK) & 0.38\% $\pm$ 0.40\% & .11\% & -.19\% & .85\% & -0.02\% & 0.66\% & 0.22\% & 0.80\% & 0.58\% \\ 
  \hline
 T$_{1}$ (\textmu s) & 45 & 53 $\pm$ 12 & 50 $\pm$ 6 & 37 $\pm$ 5 & 56 $\pm$ 10 & 55 $\pm$ 9 & 35 $\pm$ 5 & 26 $\pm$ 2 & 52 $\pm$ 7\\ 
  \hline
 T$_{2}^{*}$ (\textmu s) & 31 & 21 $\pm$ 4 & 60 $\pm$ 15 & 29 $\pm$ 6 & 33 $\pm$ 5 & 23 $\pm$ 6 & 27 $\pm$ 5 & 31 $\pm$ 3 & 26 $\pm$ 4\\ 
   \hline
 T$_{2Echo}$ (\textmu s) & 49 & 42 $\pm$ 5 & 71 $\pm$ 9 & 55 $\pm$ 6 & 47 $\pm$ 5 & 49 $\pm$ 5 & 51 $\pm$ 7 & 38 $\pm$ 2 & 37 $\pm$ 3 \\ 
  \hline
\end{tabular}}
\caption {\label{tab:coh}Qubit properties from devices fabricated with the high-uniformity junction recipe of wafers 35, 37, and 38. Frequency estimates from room temperature junction resistance and simulated capacitance show good agreement with actual qubit frequencies, especially when considering detunings between neighbors. Coherences are quoted as the average value $\pm$ 1 standard deviation of the observed temporal fluctuations during 50 hours of data acquisition.}
\end{table*}
% \normalsize

Sample \#1 was cooled to 8 mK in a BluFors XLD400 with a mixing chamber shield, cryoperm shield, Sn shield, and an inner radiation shield. The sample box was light tight and In sealed. All 8 control lines and common readout bus were connected to coaxial lines from room temperature that were attenuated (60 dB for control lines, 80 dB for readout) and low pass filtered using both LC and lossy Eccosorb filters. Roughly 60 dB of isolation was installed between the sample and the HEMT. Sample \#2 was cooled to 14 mK  in an Oxford DR200 with a cyroperm shield, inner radiation shield, and a previous generation In sealed sample box. Due to the limited number of coaxial lines in this refrigerator, qubit control pulses were injected through the common readout bus and the control lines were wirebonded to the PCB and 50 \textOmega~terminated on the cryopackage. The readout line had 80 dB of attenuation, LC and Eccosorb low pass filters, and 80 dB of isolation from the HEMT.

 The qubit energy relaxation times, T$_{1}$, dephasing times measured with a Ramsey experiment, T$_{2}^{*}$, and dephasing time found by echoing away low frequency noise with a $\pi$ pulse in the middle of the Ramsey evolution, T$_{2Echo}$, are summarized in the Table~\ref{tab:coh}. Since careful A/B testing of the box designs is still ongoing, we can only suggest possible mechanisms for the reduced coherence of sample \#2 compared to sample \#1. We suspect that T$_{2}^{*}$ and T$_{2Echo}$ may have been improved if the 50 \textOmega~terminations on control lines were replaced with shorts to reduce shot noise dephasing. Additionally, sample \#2 was characterized at a higher temperature in a fridge with less shielding.  
 
 Sample \#1 had larger frequency offsets between room temperature probing estimates and cryogenic measurements than sample \#2. As mentioned above, in preparation for cooldown, samples are cleaned in 1165 and then probed. Sample \#1 spent 24 hours at room temperature after probing (with $\approx$ 20 of those hours under vacuum) whereas sample \#2 only spent 2 hours at room temperature. If the 1165 induced aging that occurs at atmosphere also occurs under vacuum, this could explain the larger frequency offset for sample \#1. However, the small standard deviation of the frequency differences for both samples ($< 0.5\%$) shows that relative detunings between qubit pairs can be accurately predicted with room temperature probing.

% \pagebreak
% 